\documentclass[prl,twocolumn,aps]{revtex4}
\usepackage{graphicx}

\begin{document}

\title{Excitation Chains at the Glass Transition}

\author{J. S. Langer}

\affiliation{Department of Physics,
University of California,
Santa Barbara, CA  93106-9530  USA}

\date{June, 2006}

\begin{abstract}
The excitation-chain theory of the glass transition, proposed in an earlier publication,  predicts diverging, super-Arrhenius relaxation times and, {\it via} a similarly diverging length scale, suggests a way of understanding the relations between dynamic and thermodynamic properties of glass-forming liquids.  I argue here that critically large excitation chains play a role roughly analogous to that played by critical clusters in the droplet model of vapor condensation. The chains necessarily induce spatial heterogeneities in the equilibrium states of glassy systems; and these heterogeneities may be related to stretched-exponential relaxation. Unlike a first-order condensation point in a vapor, the glass transition is not a conventional phase transformation, and may not be a thermodynamic transition at all.
\end{abstract}

\maketitle

In \cite{JSL06} (see also \cite{LL05}), I proposed that both the dynamics and thermodynamics of glass-forming liquids might be determined by chainlike excitations of intrinsically disordered molecular configurations.  Because these excitations appeared to pertain only to irreversible behavior -- {\it e.g.} to anomalously slow, super-Arrhenius, viscous relaxation -- it seemed remarkable that they could also govern equilibrium properties; thus I speculated that this situation might imply partial violation of ergodicity at temperatures above the glass transition.  After rethinking this situation, I now believe that such a drastic departure from conventional statistical mechanics is unnecessary.  Rather, I think that the excitation chains allow us to understand the transition between a glass-forming liquid and a glass as an unconventional kind of phase transformation or, quite possibly, not as a thermodynamic phase transformation at all.

The most successful descriptions of the entropic collapse of glass forming liquids near the Kauzmann temperature have been spin-glass theories in which random, infinitely long-range interactions between Ising-like spins simulate glassy disorder.  Such systems do not look much like liquids with short-range interactions between the molecules.  Moreover, although such models produce qualitatively realistic thermodynamic behavior, it has been difficult to use them to compute super-Arrhenius dynamics.  Bouchaud and Biroli \cite{BOUCHAUD04}, in reviewing this situation, have made an especially thoughtful effort to find a connection between thermodynamics and dynamics {\it via} the mean-field theories.  In my opinion, their analysis reveals deep problems associated with any attempt to describe localized phenomena such as droplets or domain boundaries in a theory where the range of interactions is as long as the largest length scales in the system.  Thus, I think it is useful to explore a different approach based only on short ranged interactions between ordinary molecules.   

Chainlike (or stringlike) excitations appear to be ubiquitous in the neighborhood of jamming transitions in amorphous materials.  For example, see the work of Glotzer {\it et al.} \cite {GLOTZER}  A more recent  example that seems relevant to the glassy state itself, just below the glass transition, is seen in Fig. 4 of Silbert {\it et al.} \cite{SILBERT}, which shows a low-frequency vibrational mode of a granular material near an un-jamming transition.  The largest displacement amplitudes lie predominantly along stringlike paths. As in \cite{JSL06}, I visualize an excitation chain in a glass as a string of molecular displacements producing the glassy equivalent of a ``vacancy'' at one end and an ``interstitial'' at the other.  That is, an excitation chain is a momentary density fluctuation that most probably lasts for only a few  molecular vibration periods, but may under some circumstances induce a long lasting change in the system, moving it from one inherent state to another.  It is the latter possibility, where the vacancy and interstitial in effect dissociate from each other, that is relevant to alpha relaxation processes above the glass transition.  These chainlike excitations also must be present in the glassy state below the transition, where they cannot enable the system to explore different inherent states but must be present and contribute -- perhaps only negligible amounts -- to the thermodynamics because they are thermal fluctuations just like any other thermal excitation of the system. 

There is an interestingly imperfect analogy between the excitation-chain model of a glass and the droplet model of a vapor in the neighborhood of its condensation point.  In the latter, one approximates the vapor as a noninteracting gas of liquid droplets.  The partial pressure associated with droplets consisting of $n$ molecules, when $n$ is large, is assumed to be $p_D(n)\propto \exp (- f_n/k_B T)$ where $f_n$ is the free energy of a size-$n$ cluster: $f_n \approx h\,n + \sigma\,n^{2/3}$. Here $h$ is the chemical potential measured from the two-phase coexistence point and $\sigma$ is proportional to the surface energy.  In the stable phase, {\it i.e.} for $h>0$, $p_D(n)$ is an exponentially decreasing function of $n$ for arbitrarily large $n$.  Any cluster of molecules that reaches a substantial size $n$ is more likely to evaporate than to grow.  On the other hand, in the metastable phase with $h<0$,  there is a critically large $n$ beyond which $p_D(n)$ begins to increase, meaning that droplets larger than this critical size are more likely to grow than to evaporate.  This critical droplet is the transition state for nucleation;  thus, historically, the droplet model has been used primarily for estimating condensation rates.  Equilibrium thermodynamic properties (away from critical points) are determined primarily by small clusters, {\it i.e.} low-order terms in the virial expansion, for which the droplet approximation makes little sense.  There also are intrinsic uncertainties in computing $f_n$ for realistic situations even for very large $n$; in particular, it is not clear how to decide which clusters of molecules are close enough together to be counted as droplets.  Such uncertainties have largely been eliminated in field-theoretic calculations by looking only in the immediate neighborhood of the transition state, {\it i.e.} at the ``instanton'' \cite{JSL67,JSL69}, and computing the flux through this point in  function space to find the condensation rate.  The instanton theory also predicts a branch cut in the analytic continuation of the free energy from the stable to the metastable phase.  The discontinuity across the cut is proportional to the condensation rate; thus there is a deep connection between equilibrium and nonequilibrium behaviors in these systems. 

The excitation-chain model of a glass has much the same structure as the droplet model of a vapor.  In \cite{JSL06}, I argued that the probability of thermal activation of an excitation chain consisting of $N$ molecular steps and extending a distance $R$ (in units of the average molecular spacing) is proportional to $\exp \bigl(-\Delta G(N,R)/k_B\,T\bigr)$, where $\Delta G(N,R)$ is a free energy in the sense that it includes a sum over all the allowed configurations of the chain.   $\Delta G(N,R)$ consists of several parts:
\begin{equation}
\label{DeltaG1}
\Delta G(N,R)= \Delta G_{\infty} + N e_0 + E_{int} - k_B\,T\,\ln W .
\end{equation}
The first term, $\Delta G_{\infty}$, is the bare activation energy, {\it i.e.} the energy required to form the vacancy and the interstitial regardless of whether or not these defects are well separated from each other.  The remaining terms on the right-hand side of Eq.(\ref{DeltaG1}) describe the excess free energy of the chain.  $N e_0$, is the bare activation energy of the $N$ links of the chain. The average energy per link, $e_0$, is the energy required to move a pair of molecules far enough away from each other to allow a third molecule to pass between them. $E_{int}$ is an energy that makes it unfavorable for the links of the chain to lie near each other.  Here I have used Flory's approximation for the self-exclusion energy of a polymer chain:
\begin{equation}
\label{Eint}
E_{int}(N,R)\approx k_B\,T_{int}\,{N^2\over R^3}.
\end{equation}

The last term on the right-hand side of Eq.(\ref{DeltaG1}), $W(N,R)$, is a sum over chain configurations.  In \cite{JSL06} I wrote $W$ in the form:
\begin{equation}
\label{W1}
 \ln\,W(N,R) \approx  \nu\,N - {\pi\,\gamma(T)\over 2}\,R,  
\end{equation}
where $\exp\,(\nu)$ is the number of choices that the successive links in the chain can make at each step, and $\gamma(T) = \gamma_0\,(T_0/T)^2$ is the mean-square strength of the fluctuations in $e_0/k_B T$.  Here, $T_0 = e_0/\nu\,k_B$ is the characteristic temperature determined by the energy $e_0$; and $\gamma_0$ is an inverse localization length associated with the fact that the excitation chains exist in the highly disordered environment characteristic of a molecular glass.  Because lengths are measured in intermolecular spacings, and glassy disorder is on molecular scales, $\gamma_0$ must be of order unity near the glass temperature.  (I omit the diffusion term, $R^2/2N$, from the usual random-walk analysis because it is negligible for present purposes.)  Note that both the Flory approximation in Eq.(\ref{Eint}) and the configuration-counting approximation in Eq.(\ref{W1}) make sense only in the limit of large $N$ and $R$. 

The next step in this analysis is to evaluate the formation probability for all chains of length $N$
by finding the value of $R = R^*$ at which $\Delta \,G(N,R)$ is a minimum, and then approximating this probability by $
p_{XC}(N) \propto \exp \left(-\Delta\,G^*(N)/k_B T\right)$,
where
\begin{eqnarray}
&&\Delta \,G^*(N) = \Delta\,G(N,R^*)= \Delta G_{\infty} - N\,k_B\,\nu\,(T-T_0)\cr && + 
4\,\left(\pi/6\right)^{3/4}\,(k_B\,T_{int})^{1/4}\,(\gamma(T)\,k_B\,T)^{3/4}\,N^{1/2}.
\end{eqnarray}
Note the similarity between $p_{XC}(N)$ and the formula for $p_D(n)$ in the droplet model. For $T < T_0$, $p_{XC}(N)$ decreases monotonically for all $N$.  However, for $T > T_0$, $\Delta \,G^*(N)$ has a maximum, and $p_{XC}(N)$ a minimum, at a value of $N$, say $N^*$, which I identified in \cite{JSL06} as the length of the critically large chain that nucleates a long-lasting  density fluctuation.  If a chain fluctuates to a size larger than $N^*$, it is highly likely to continue growing without bound, and thus to dissociate the vacancy-interstitial pair.  The activation energy for this process is
\begin{equation}
\label{DeltaGnew}
{\Delta G^*(T)\over k_B T} = {\Delta G(N^*,R^*)\over k_B T}= {\Delta G_{\infty}\over k_B T}+ {\pi\over 3}\,\gamma(T)\,R^*(T),
\end{equation}
where the length $R^*(T)$ is
\begin{equation}
\label{R*}
R^*(T)= 2\,\left({\pi\over 6}\right)^{1/2}{\gamma(T)^{1/2}\,(T\,T_{int})^{1/2}\over \nu\,(T-T_0)}.
\end{equation}
Equations (\ref{DeltaGnew}) and (\ref{R*}) recover the Vogel-Fulcher law for the logarithm of the $\alpha$ relaxation time $\tau_{\alpha}$, which diverges linearly at $T_0$.  Thus the excitation-chain mechanism provides a simple explanation for super-Arrhenius behavior in glasses, and also predicts a diverging length scale $R^*(T)$.  

My thesis is that the excitation chains play a role near the glass transition analogous to that played by large droplets near a condensation point.  The equilibrium partition sum in the glass phase, below $T_0$, includes a dilute population of long excitation chains that, like the large droplets, contribute negligibly to the thermodynamics but are present nonetheless.  As in the droplet model, the overwhelmingly most important contributions to the thermodynamics come from small, localized, thermal displacements of the molecules, which may sometimes resemble short chains in a glass.  We certainly do not want to use the long-chain formulas to estimate the contributions of the short chains.  Just as in the droplet model where we cannot decide which molecular clusters are to be counted as droplets, there is no unambiguous definition of a thermally excited chain; we do not know how large the molecular displacements along the chain must be in order for a sequence of such displacements to be counted as a ``chain.''  The best we can do is guess that these displacements are of order the molecular spacing for critical chains of size $N^*$, and then suppose that this guess will become more precise if we can construct the analog of an instanton theory for this system. 

The crucial point is that large, rare, excitation chains are intrinsically part of the equilibrium state of a glassy system, both below and above $T_0$, and that they play an especially important role above $T_0$ where they nucleate transitions between inherent states.  With this understanding, we may deduce that the thermal equilibrium state of a glass-forming liquid is necessarily heterogeneous on a length scale of order $R^*(T)$.  A homogeneous glassy region much larger than $R^*(T)$, according to the formula for $p_{XC}(N)$, would contain a substantial population of chains longer than $N^*$ and thus would be unstable against changes in size and shape.  Conversely, an isolated domain smaller than $R^*$ could not support excitation chains and therefore would be immobile.  Presumably, the molecules in the interiors of these domains are frozen into energetically favorable glassy configurations determined by their frustrated, short-range interactions -- just as in the glassy state below $T_0$.  The boundaries between these domains must then consist of more disordered material.  In the language I have used here, they contain a high density of vacancies and interstitials.  Pictures of such a domain structure have appeared in numerical simulations.  For example, see Figs. 6b and 7 in \cite{YAMAMOTO-ONUKI}.

Note that the excitation chains cannot cross the disordered boundaries, where the vacancies and interstitials at the ends of the chains can recombine with preexisting interstitials and vacancies.  These recombinations cause the domain boundaries to wander slowly on the time scale $\tau_{\alpha}$.  As a result of this wandering, there is no breaking of ergodicity above $T_0$. The system samples all of its state space, but does so increasingly slowly as $T$ decreases toward $T_0$. The glass-forming liquid freezes continuously as $T$ decreases; the frozen domains become larger and the unfrozen molecules in the boundaries between the domains become a vanishingly small fraction of the system.  Over long times at fixed $T$, any individual molecule is sometimes in a frozen state (inside a domain, locked into a cage by its neighbors, and contributing to the entropy as if it were in the glassy state below $T_0$ ), and is sometimes unfrozen (in a boundary region, contributing to the entropy as if it were more nearly in a liquid state).  At any given time, the unfrozen fraction of the molecules is proportional to the surface-to-volume ratio of the domains, $3/R^*(T)$.  Thus the  configurational entropy per molecule, {\it i.e.} the excess over the residual entropy in the frozen glass, in units of $k_B$, is
\begin{equation}
\label{entropy}
s_c(T)\approx {3\,s_0\over R^*(T)} = {3\over 2}\,\left({6\over \pi}\right)^{1/2}\,{s_0\,\nu\,(T-T_0)\over (\gamma_0\,T_0\,T_{int})^{1/2}},
\end{equation}
where $s_0$ is the configurational entropy per unfrozen molecule, multiplied by the molecular thickness of the inter-domain boundary, which I guess is independent of the size of the domains. 

Equation (\ref{entropy}) implies that the excess configurational entropy vanishes linearly in $T$ at the Kauzmann temperature $T_K$, and that $T_K = T_0$.  The combination of Eqs.(\ref{DeltaGnew}) and (\ref{entropy}) yields
\begin{equation}
{\Delta G^*(T)\over k_B\,T} \approx {\pi\,\gamma_0\,s_0\over s_c(T)},
\end{equation}
which is the Adam-Gibbs formula. \cite{ADAM-GIBBS65}  

Definitions of the glass temperature $T_g$ generally have the form $\Delta G^*(T_g)/ k_B T_g  = \lambda_g$, where $\lambda_g$ is a number of order $30$, chosen roughly to represent the observable limits of long relaxation times or high viscosities.  Then, because $\Delta G^*(T_g)$ diverges near $T_0$, $T_g\cong T_0$, the fragility $m$ is \cite{ANGELL95}  
\begin{equation}
m \equiv - T{\partial\over \partial T}\left({\Delta G^*(T)\over k_B\,T}\right)_{T=T_g} \propto {\nu\,\lambda_g^2\over \gamma_0}\left({T_g\over \gamma_0 T_{int}}\right)^{1/2}
\end{equation} 
Returning to the thermodynamic formula, Eq.(\ref{entropy}), we find that the jump in the specific heat at $T_g \cong T_0$ is
\begin{equation}
\label{Deltacp}
\Delta c_p = \left(T\,{\partial s_c\over \partial T}\right)_{T=T_0} \approx {6\,s_0\,\gamma_0\,m\over \lambda_g^2}.
\end{equation}
Here we recover the conjectured proportionality between $\Delta c_p$ and $m$, but with two material-specific parameters that might account for the observed scatter in the experimental data.  $s_0$ should scale with the ``bead'' number of the molecules, which ordinarily is factored out in obtaining the linear relation in Eq.(\ref{Deltacp}).  \cite{WOLYNES00}  Because it is basically a geometrical quantity, $\gamma_0$ may be roughly a constant, of order unity.  This analysis also implies that
\begin{equation}
R^*(T_g)\approx \left({3\over \pi}\right)\,{\lambda_g\over \gamma_0}.
\end{equation}
Thus the critical length scale $R^*$ at the glass temperature is predicted to be independent of the fragility, in qualitative agreement with results shown by Berthier {\it et al.} \cite{BERTHIER05}; but those authors report a substantially smaller length scale.  

The droplet-model analogy suggests that some kind of transformation is occurring at $T_0$, but it is not clear whether this is a true phase transition or, if so, what kind of transition it might be.  Dissociation of a vacancy-interstitial pair does not nucleate a qualitatively new state of the whole system, as does the appearance of a critically large droplet in a supersaturated vapor; it is simply a mechanism by which the glass takes steps in exploring its configuration space and thus  starts to behave like a liquid. As discussed in \cite{JSL06}, there is a  super-Arrhenius range of temperatures, $T_0 < T < T_A$, within which the excitation-chain mechanism is operative and the $\alpha$ relaxation time has approximately the Vogel-Fulcher temperature dependence.  $T_A$ is the temperature at which $N^*$ and $R^*$ become small, {\it i.e.} where excitation chains are no longer needed to stabilize density fluctuations.  Above $T_A$, relaxation is Arrhenius -- the activation energy is temperature independent -- and the system behaves like an ordinary, unjammed liquid, albeit like one where molecular rearrangements still must overcome activation barriers.  The transition from the glass to the liquid occurs continuously across the super-Arrhenius region, with no latent heat, nor any change in symmetry.  

The excitation-chain picture does not require that $T_0$ or $T_A$ be sharply defined temperatures.  The divergence of $\Delta G^*(T)$ at $T=T_0$ in Eq. (\ref{DeltaGnew}) may be an artifact of the approximations made in deriving Eq.(\ref{W1}). (See \cite{JSL06} for a discussion of the uncertainties at low $T$ and large $N$, where the discreteness of the glassy molecular structure must become increasingly relevant.) The actual relaxation rates may remain finite but unobservably small down to $T=0$, in which case there would be no thermodynamic phase transition at all. In principle, ergodicity would be restored at all temperatures, but thermodynamic equilibrium well below $T_g$ would be experimentally inaccessible.  Such a possibility would be consistent with the recent results of Donev {\it et al.} \cite{DONEV06}, who find that a binary mixture of hard disks cannot exhibit a  transition to an ideal glassy state of zero entropy, although that system does behave in a glasslike manner at sufficiently high densities.  

Similarly, we know that the large-$N$ approximations used in Eqs. (\ref{Eint}) and (\ref{W1}) are invalid for small $N$ and cannot produce the smooth transition between super-Arrhenius and Arrhenius behaviors that is seen experimentally at $T_A$. In all likelihood, $T_A$ has no special thermodynamic significance, although it does mark the approximate upper bound of the temperature range in which heterogeneities on the scale $R^*(T)$ are present.  

The spatial and temporal heterogeneities implied by this theoretical picture make it likely that a closer look at relaxation processes will reveal stretched exponential decays of correlations.  A detailed discussion of this topic is beyond the scope of the present report,  but one illustration may be useful.  (The following analysis has some features in common with the trapping models that have been discussed, for example, in references \cite{G-P,PHILLIPS,BENDLER02}).   For $T$ just above $T_0$, any individual molecule spends most of its time immobilized within a frozen domain.  Occasionally, a boundary region wanders through its position and allows it to diffuse a substantial distance before being frozen again.  If the molecule is frozen in a domain of size $R$, the probability of its remaining in place for a time $t$ (in units of $\tau_{\alpha}$) is proportional to $\exp (-t/R^2)$. Then, if the distribution of domain sizes is Gaussian with a width $R^*$, $\sim \exp [-(R/R^*)^2]$, the most likely value of $R$ scales like $t^{1/4}$, and the relaxation function becomes $\exp (- 2 t^{1/2}/R^*)$.  If this crude estimate is valid, the limiting value of the stretching exponent would be $1/2$.  As $T$ increases toward $T_A$, the domain size decreases and the exponent must return to unity in a smooth way that -- like other features of the transition between super-Arrhenius and Arrhenius behavior -- remains to be determined in this theory.   

In summary, I believe that the excitation-chain hypothesis may provide a framework -- but not yet more than a framework -- for constructing a theory of the glass transition based on realistic  molecular models.  Outstanding challenges include improved estimates of activation rates for both long and short chains, a systematic statistical mechanical derivation of the entropy formula in Eq.(\ref{entropy}), and an analysis of how and where stretched-exponential relaxation might emerge from this model.  

\begin{acknowledgments}
This research was supported by U.S. Department of Energy Grant No. DE-FG03-99ER45762. I would like to thank  Jean-Philippe Bouchaud for very helpful discussions.
\end{acknowledgments}

\end{document}